\documentclass{article}

\usepackage{amsmath,amssymb} 
\usepackage[]{graphicx}
\graphicspath{{graphics/}} 
\usepackage{color}
\usepackage[linktocpage]{hyperref}
\usepackage{wrapfig}
\usepackage{ulem}
\usepackage{lscape}

\def\eps{\varepsilon}
\newcommand{\fraz}{\displaystyle\frac}
\def\##1{{\bf #1}}
\def\=#1{\underline{\underline #1}}
\def\.{\mbox{ \tiny{$^\bullet$} }}
\def\ux{\#u_x}
\def\uy{\#u_y}
\def\uz{\#u_z}

\def\ped0{_{\scriptscriptstyle 0}}
\def\ko{k\ped0}
\def\tq{\tilde{q}}
\def\vph{v_{ph}}
\def\propdist{\Delta_{prop}}
\def\fref#1{Figure~\ref{#1}}

\def\sref#1{Section~\ref{#1}}
\def\tref#1{Table~\ref{#1}}

\def\tond#1{\left(#1\right)}
\def\quadr#1{\left[#1\right]}

\begin{document}

\begin{center}

{\large Compound surface-plasmon-polariton waves guided by a thin metal
layer sandwiched between a homogeneous isotropic dielectric
material and a structurally chiral material} \vskip 0.4cm

{\bf Francesco Chiadini}$^1$, {\bf Vincenzo Fiumara}$^2$, {\bf Antonio Scaglione}$^2$, {\bf Akhlesh Lakhtakia}$^3$

\vskip 0.2cm $^1${\textit Department of Industrial Engineering,
University of Salerno,\\via Giovanni Paolo II, 132 - Fisciano (SA), 
84084, ITALY\\
e-mail: fchiadini@unisa.it, ascaglione@unisa.it}

$^2${\textit School of Engineering, University of 
Basilicata,\\Viale 
dell'Ateneo Lucano 10, 85100 Potenza, ITALY\\ email: vfiumara@unibas.it}

$^3${\textit Department of Engineering Science and Mechanics, Pennsylvania State University,\\
University Park, PA 16802--6812,
USA\\
e-mail: akhlesh@psu.edu}

\end{center}

\vskip 0.5cm



\begin{abstract}
Multiple compound surface plasmon-polariton (SPP)  waves  can be guided by a structure consisting of  a sufficiently thick layer of
metal  sandwiched between a homogeneous isotropic dielectric (HID) material and a dielectric structurally chiral material  (SCM).  The compound SPP waves are strongly bound to both metal/dielectric interfaces when  the thickness of the metal layer is comparable to the skin depth but just to one of the two interfaces when the thickness is much larger. The compound SPP waves differ in phase speed, attenuation rate, and field profile, even though all are excitable at the same frequency.  
Some compound SPP waves are not greatly affected by the choice of the direction of propagation in the transverse plane but others are, depending on metal thickness.
For fixed metal thickness, the number of compound SPP waves depends on the relative permittivity of the HID material, which can be useful for sensing applications.

\end{abstract}




\section{Introduction}
\label{sect:intro} 
The propagation of a surface-plasmon-polariton (SPP) wave is guided by a planar metal/dielectric  interface, the field strengths decaying exponentially away from the interface in both materials~\cite{Pitarke}. If the dielectric material is homogeneous and isotropic, only one  SPP wave can propagate parallel to the interface at a specific  frequency. 
If the dielectric material is  periodically nonhomogeneous in the direction normal to the interface, multiple SPP waves that differ in polarization state, phase speed, attenuation rate, and field profile, can be guided simultaneously by the interface at a specific  frequency~\cite{AkhBook}. 

This multiplicity is attractive for optical sensing applications, as the sensitivity and reliability of sensing can be enhanced thereby \cite{SPL}. Furthermore, the 
number of the simultaneously detected analytes can than also be greater than one, the usual number \cite{HomolaBook}. The
same multiplicity will enhance SPP-wave-based microscopy \cite{Stabler} and communications \cite{Sekhon} as well.

For about three decades, one way to further increase the number of SPP waves is to interpose a thin metal layer between two suitably chosen homogeneous dielectric materials \cite{Sarid,QRS1983}. The two SPP waves, each guided separately by  a metal/dielectric interface when the metal layer is thick, hybridize into compound SPP waves.
Motivated by that possibility,
recently we analyzed the propagation of multiple compound SPP waves guided by an isotropic metal layer sandwiched between a homogeneous isotropic dielectric (HID) material and a periodically multilayered isotropic dielectric (PMLID) material. We demonstrated that compounding occurs even when one of the two metal/dielectric interfaces is capable of guiding multiple SPP waves by itself~\cite{NoiJNP}.

In this paper, we extend the scope of the multiple-compound-SPP-wave phenomenon to encompass anisotropic dielectric materials by
replacing the PMLID material by a dielectric structurally chiral material (SCM). Exemplified by Reusch piles \cite{Reusch,JB2}, cholesteric
liquid crystals   \cite{Chandra,deGennes}, and chiral sculptured thin films 
\cite{AkhNatSTF}, a dielectric SCM is anisotropic and
helically nonhomogeneous along a fixed axis. Whereas a planar metal/HID interface by itself can guide a single SPP wave, the planar interface of a metal and a dielectric SCM
 by itself can guide multiple SPP waves \cite{PLprsa}.

This paper is organized as follows. A description of the boundary-value problem is provided
 in \sref{sec:theo}, but we have elected not to describe the   procedure to obtain the dispersion equation
 for compound SPP waves, as the  methodology is
 available in detail elsewhere \cite[Chap.~3]{AkhBook}. In \sref{sec:numres} numerical results showing the compounding of the SPP waves guided by the metal/HID and metal/SCM interfaces are presented in relation to both the thickness of the metal layer, the relative permittivity of the HID,  and the direction of propagation in the transverse plane. Concluding remarks are given in \sref{sec:concl}. 

 An $\exp\tond{-i\omega t}$ dependence on time $t$ is implicit, with $\omega$ denoting the angular frequency and $i=\sqrt{-1}$. Furthermore, $\ko =\omega \sqrt{\eps\ped0\mu\ped0}$  and $\lambda\ped0=2\pi/\ko$, respectively, represent the wavenumber and the wavelength
 in free space, with $\mu\ped0$ as the permeability and $\eps\ped0$ as the permittivity of free space. Vectors are in boldface;   dyadics are double underlined;   Cartesian unit vectors are denoted by $\ux$, $\uy$, and $\uz$; and the asterisk denotes the complex conjugate.

 \begin{figure}
 \begin{center}
 \begin{tabular}{c}
\includegraphics[width=0.9\linewidth]{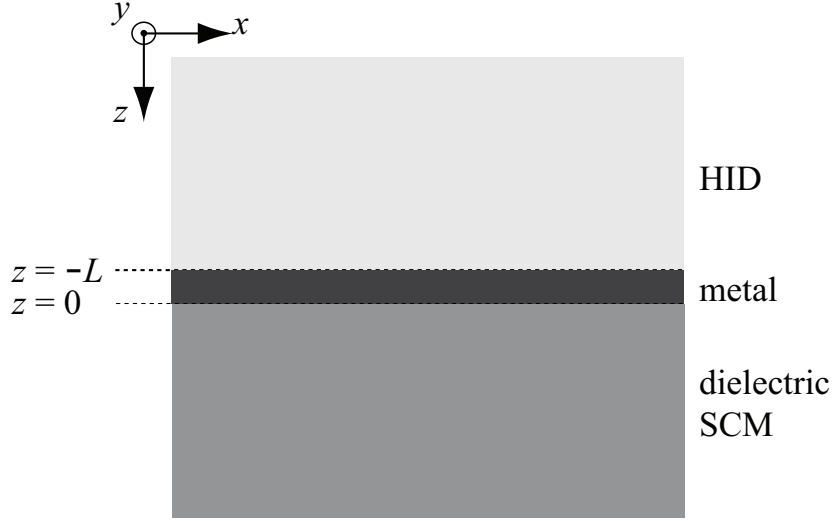}
 \end{tabular}
 \end{center}
\caption{Schematic of the boundary-value problem. A metal layer of thickness $L$ separates a half space
occupied by a HID material and another half space occupied by a dielectric SCM.}
\label{fig:schem}
 \end{figure} 

\section{Theoretical framework}\label{sec:theo}

The geometry of the boundary-value problem for the propagation of compound SPP waves is schematically illustrated in \fref{fig:schem}. The half space $z < -L$ is occupied by a HID material with real-valued relative permittivity $\eps_d>0$. An isotropic metal layer of thickness $L$ and complex-valued relative permittivity $\eps_m$ separates the HID material from a
dielectric SCM which occupies the half space $z>0$. The chosen SCM  is periodically nonhomogeneous along the $ z $ axis with period $ 2\Omega $ and can be characterized by the relative-permittivity dyadic~\cite{AkhBook} 
\begin{equation}
\=\eps_{SCM}\left(z\right) = \=S_z\left(h\pi{z}/\Omega\right)\. \=S_y\left(\chi \right) \. \=\eps_{ref}^o \. \=S_y^{-1}\left(\chi \right)\. \=S_z^{-1}\left(h\pi{z}/\Omega\right)\,,
\label{epschiral}
\end{equation}
where $ h =1 $ for a  right-handed SCM or $h=-1$ for a left-handed SCM; the dyadic
\begin{equation}
\=\eps_{ref}^o= \eps_a\uz \uz + \eps_b\ux\ux + \eps_c \uy \uy 
\label{epsref}
\end{equation}
contains the principal relative permittivity scalars $ \eps_a $, $ \eps_b $, and $ \eps_c $; and the rotation dyadics
\begin{equation}
\begin{aligned}
\=S_z\tond{\zeta}= \uz\uz + \left(\ux \ux\right. & \left. + \uy \uy\right)\cos \zeta  \\ &+\left(\uy \ux- \ux \uy\right)\sin \zeta
\end{aligned}
\label{Sz}
\end{equation}
and
\begin{equation}
\begin{aligned}
\=S_{y}\left(\chi \right)=  \uy \uy + \left(\ux \ux+\right.& \left.\uz \uz\right)\cos  \chi   \\ &+ \left(\uz \ux- \ux \uz\right)\sin \chi 
\end{aligned}\label{Sy}
\end{equation}
capture the helical nonhomogeneity. Whereas $\eps_a=\eps_c\ne\eps_b$ and $\chi=0$ for cholesteric liquid crystals, $\eps_a\ne\eps_b\ne\eps_c$
and $\chi\in(0,\pi/2]$ for chiral smectic liquid crystals \cite{ABW} and chiral sculptured thin films \cite{Akhbook2}. All three materials are assumed to be nonmagnetic.

Without loss of generality, we consider a compound SPP wave propagating parallel to the unit vector $\ux\cos\psi+\uy\sin\psi$,
$\psi\in[0^\circ,360^\circ)$, in the transverse (i.e., $xy$) plane  and decaying far away from the metal layer. The electric and magnetic field phasors can be written everywhere as
\begin{equation}
\begin{aligned}
\hspace{-20pt}
\left.\begin{array}{l}
\#E(x,y,z)= \left[e_x(z)\ux+e_y(z)\uy+e_z(z)\uz \right]\cdot\\
\hspace{100pt}\cdot\exp[{iq}(x\cos\psi+ y\sin\psi)]\\[5pt]
\#H(x,y,z)= \left[h_x(z)\ux+h_y(z)\uy+h_z(z)\uz \right]\cdot\\
\hspace{100pt}\cdot\exp[{iq}(x\cos\psi+ y\sin\psi)]
\end{array}\right\}\,, 
\\
\qquad\qquad\quad z \in(-\infty,\infty)\,,
\label{eq:EH1}
\end{aligned}
\end{equation}
where $q$ is  the complex-valued wavenumber.  The procedure to obtain a dispersion equation for the wavenumber $q$
is provided in detail elsewhere \cite{AkhBook}. Once $q$ has been numerically determined from that dispersion equation,
the piecewise determination of the functions $e_{x,y,z}(z)$ and $h_{x,y,z}(z)$ is also possible. 
 
\section{Numerical results}\label{sec:numres}
In this section, we present the normalized solutions $\tq =q/k\ped0 $ of the dispersion equation for compound SPP waves. The equation was  solved numerically using  Mathematica (Version 10) on a Windows XP laptop computer. For all calculations, we fixed $ \lambda\ped0=633$~nm. The HID material was taken to be SF11 glass ($ \eps_d = 3.1634 $) for the results presented in  Secs.~\ref{variableL} and \ref{sec:psi}, and a lossless dielectric material with a relative permittivity $\eps_d\in[1,4]$ for the results presented in Sec.~\ref{variablepsd}. The metal was taken to be silver ($ \eps_m =-14.4610+i1.1936$) with  skin depth $ \delta_m= 1/{\rm Im}\left(\ko\sqrt{\eps_m}\right)=26.47$~nm at the chosen wavelength. The dielectric SCM was supposed to be a chiral sculptured thin film made of patinal titanium oxide with $ \Omega = 135 $ nm, $ h=1 $, $ \eps_a=2.13952 $, $ \eps_b=3.66907 $, $ \eps_c = 2.82571 $ and $ \chi = 37.6745^{\circ} $ \cite{HWH}.  The angle $\psi$ was set equal to $0^\circ$ for the results presented in Secs.~\ref{variableL} and  \ref{variablepsd}. Results for $\psi\neq0^\circ$ are presented in the Sec.~\ref{sec:psi}.
The search for solutions was restricted to  $0<{\rm Re}(\tq) \leq 5.5$.

Once a solution $q$ of the dispersion equation was found, we determined the field phasors $\#E(\#r)$ and $\#H(\#r)$, and then
computed the Cartesian components of the time-averaged Poynting vector 
\begin{equation}
\#P(x,y,z)=\fraz{1}{2} {\rm Re} \left[\textbf{E}(x,y,z) \times \textbf{H}^\ast(x,y,z)\right]\,.
\label{TAPV}
\end{equation}
Profiles of the Cartesian components of $\#P(x,y,z)$ are presented
for representative solutions in this section.

\subsection{Effect of the thickness of the silver layer}\label{variableL}
\begin{figure}
 \begin{center}
 \begin{tabular}{c}
\includegraphics[width=0.9\linewidth]{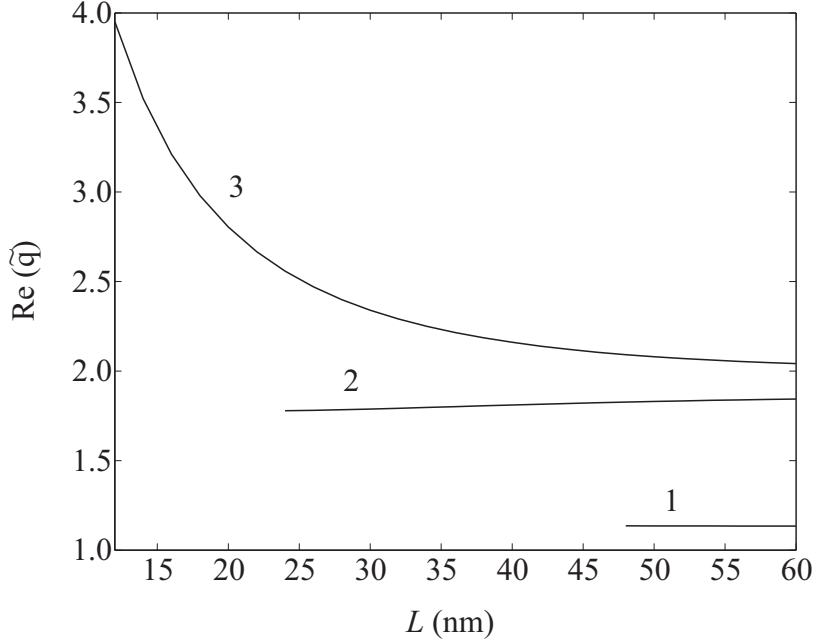}
\end{tabular}
 \end{center}
\caption{Variation of ${\rm Re}(\tq)$ with the thickness $L$ of the metal layer when $ \eps_d=3.1634 $ and $\psi=0^\circ$. 
\label{fig:ReqL}}
 \end{figure}

\begin{figure}
 \begin{center}
 \begin{tabular}{c}
	 \includegraphics[width=0.9\linewidth]{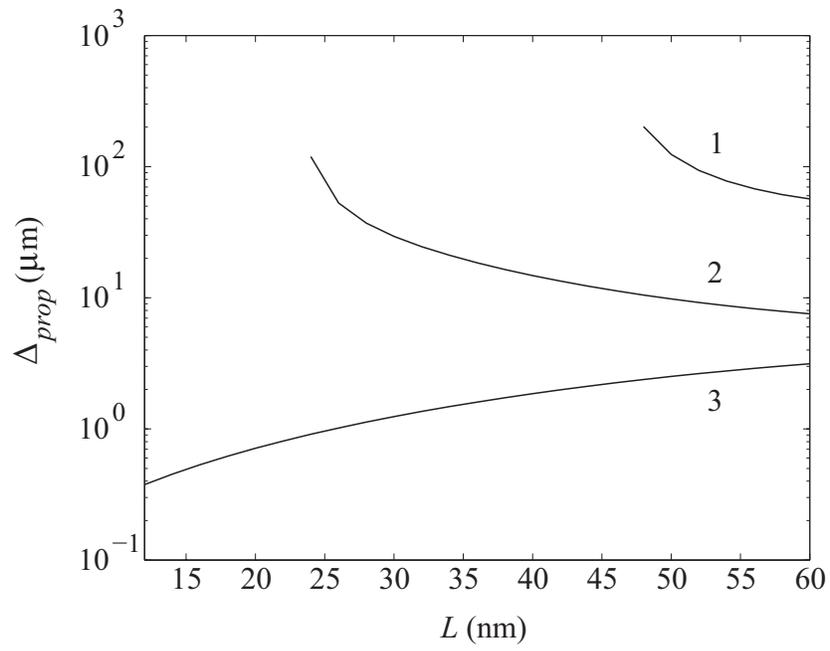}
 \end{tabular}
 \end{center}
\caption{Variation of the propagation distance $\propdist$ with the thickness $L$ of the metal layer when $ \eps_d=3.1634 $ and $\psi=0^\circ$. 
 \label{fig:DeltapropL}}
 \end{figure}

Figures \ref{fig:ReqL} and \ref{fig:DeltapropL}, respectively, show the real part of the normalized wavenumber $\tq$ and the propagation distance $\propdist=1/{\rm Im}(q)$ in the $xy$ plane found for all solutions as the thickness $ L$   of the metal layer
was varied from 10~nm to 60~nm, with $\psi=0^\circ$ fixed. As many as three different compound SPP waves can be guided by the chosen HID/metal/SCM structure for $ L\in[10,60]$~nm. These solutions are organized in three branches labeled $ 1 $ to $ 3 $ in
Figs.~\ref{fig:ReqL} and \ref{fig:DeltapropL}. Each branch spans the range $[L_{th}, 60~{\rm nm}]$, the branch being absent for $L<L_{th}$. We determined that  $ L_{th}\simeq 44 $~nm for branch~1, $ L_{th} \simeq 24 $~nm
for branch~2,  and $ L_{th}\simeq12 $~nm for branch~3. 

As $ L $ increases from $L_{th}$,  $ {\rm Re}\left(\tq\right) $ and, therefore, the phase speed $ \vph=c\ped0/ {\rm Re}\left(\tq\right)$,  are practically constant
on branches 1 and 2; however,  the propagation distance $\propdist$ decreases. Branch $3 $ has  quite different characteristics:  both $ \vph$ and $\propdist$ increase as $ L $ increases. However, for a fixed value $ L>44$~nm, the solution belonging to branch $ 1 $ exhibits the highest value of $\propdist$.

\tref{tab:Ssol} shows the relative wavenumbers $ \tq $ of the compound SPP waves computed for two different thicknesses of the metal layer, $L=60$~nm ($ > 2 \delta_m $) and $ L \sim L_{th} $, while $\psi=0^\circ$ is fixed. The table also lists the values of $\tq$ for (a) the SPP waves guided by the metal/SCM interface by itself \cite{PLprsa} and (b) the sole $p$-polarized SPP wave guided by the metal/HID interface by itself \cite{HomolaBook}. When $L=60$~nm, the relative wavenumbers of the compound SPP waves belonging to the branches $ 1 $ and $ 2 $ are very close to the ones of the SPP wave guided by the metal/SCM interface by itself, while the relative wavenumber of the compound SPP wave belonging to the branch $ 3$ is very close to the one of the SPP wave guided by the metal/HID interface alone. Such affinities indicate that, for this value of $L$, the compound SPP waves on branches $ 1 $ and $ 2 $ propagate bound predominantly to the metal/SCM interface while the compound SPP wave on branch $ 3 $ propagates bound predominantly to the metal/HID interface. Both situations are indicative of a weak coupling between the two metal/dielectric interfaces $z = -L$ and $z=0$. As $L$ decreases, solutions on branch 1 remain strongly bound to the metal/SCM interface, but solutions
on branches 2 and 3 show stronger coupling between the two metal/dielectric interfaces.

\begin{figure*}
 \begin{center}
 \begin{tabular}{c}
(a) \includegraphics[width=0.4\linewidth]{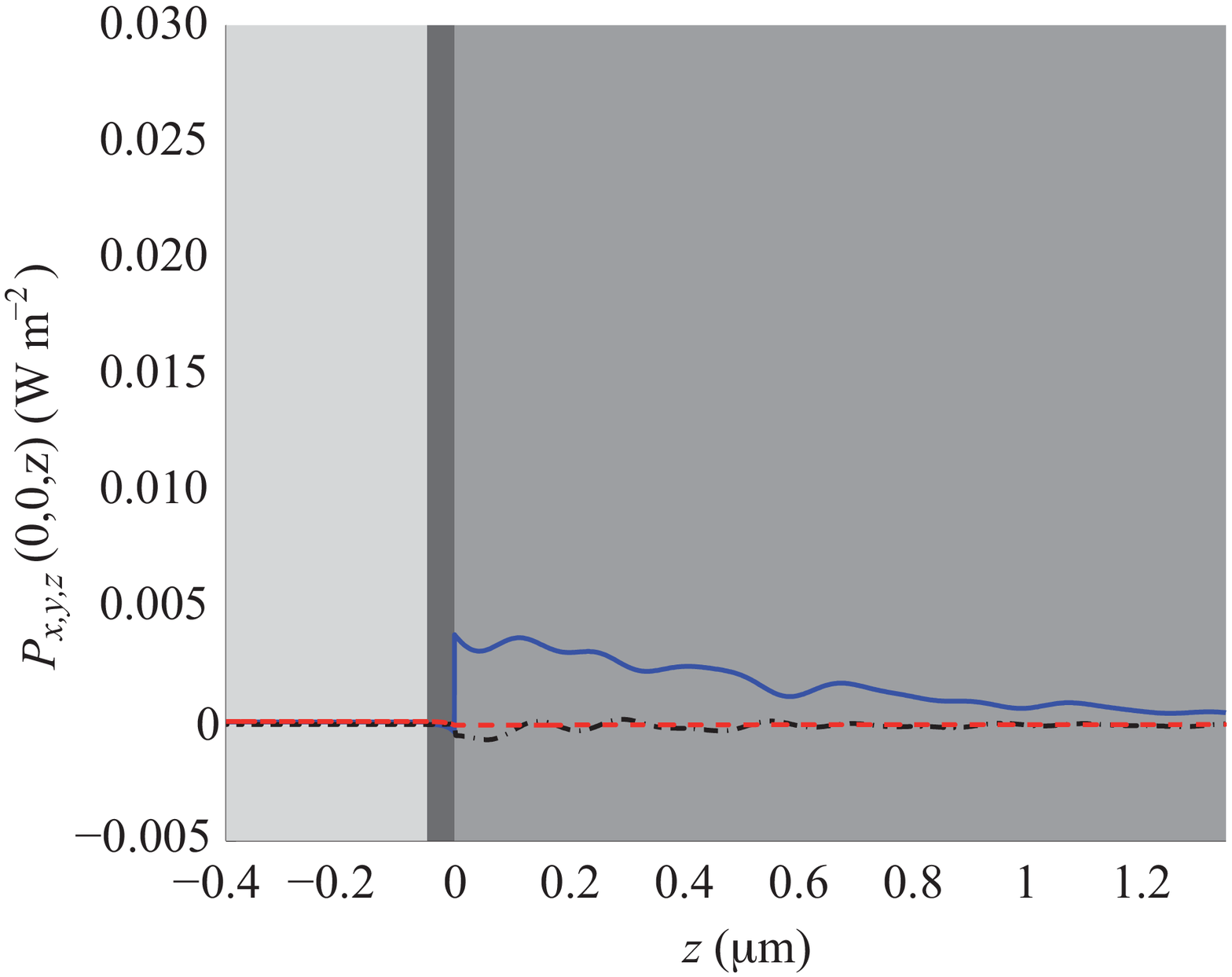}\hspace{10pt}
(b) \includegraphics[width=0.4\linewidth]{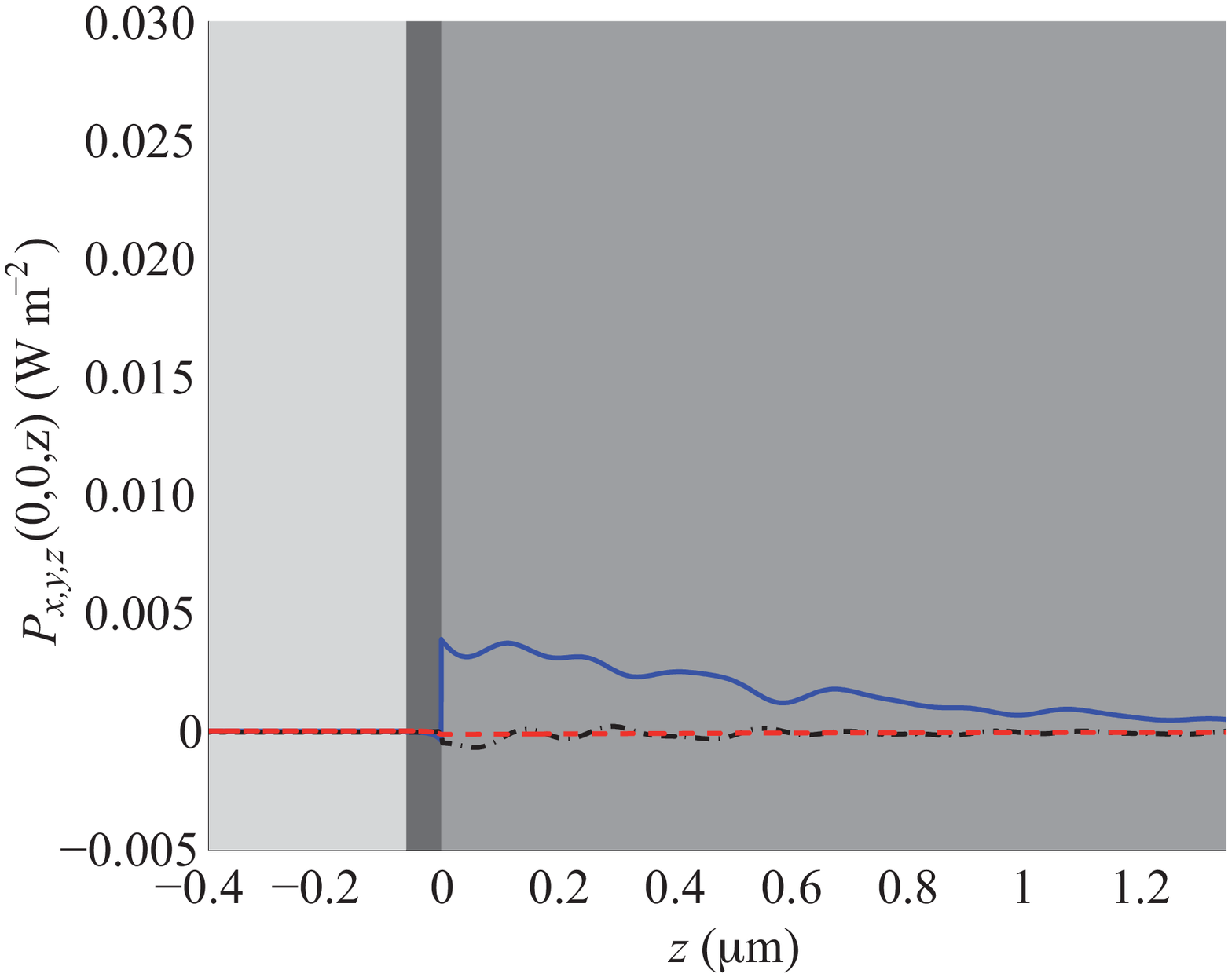}\\
(c) \includegraphics[width=0.4\linewidth]{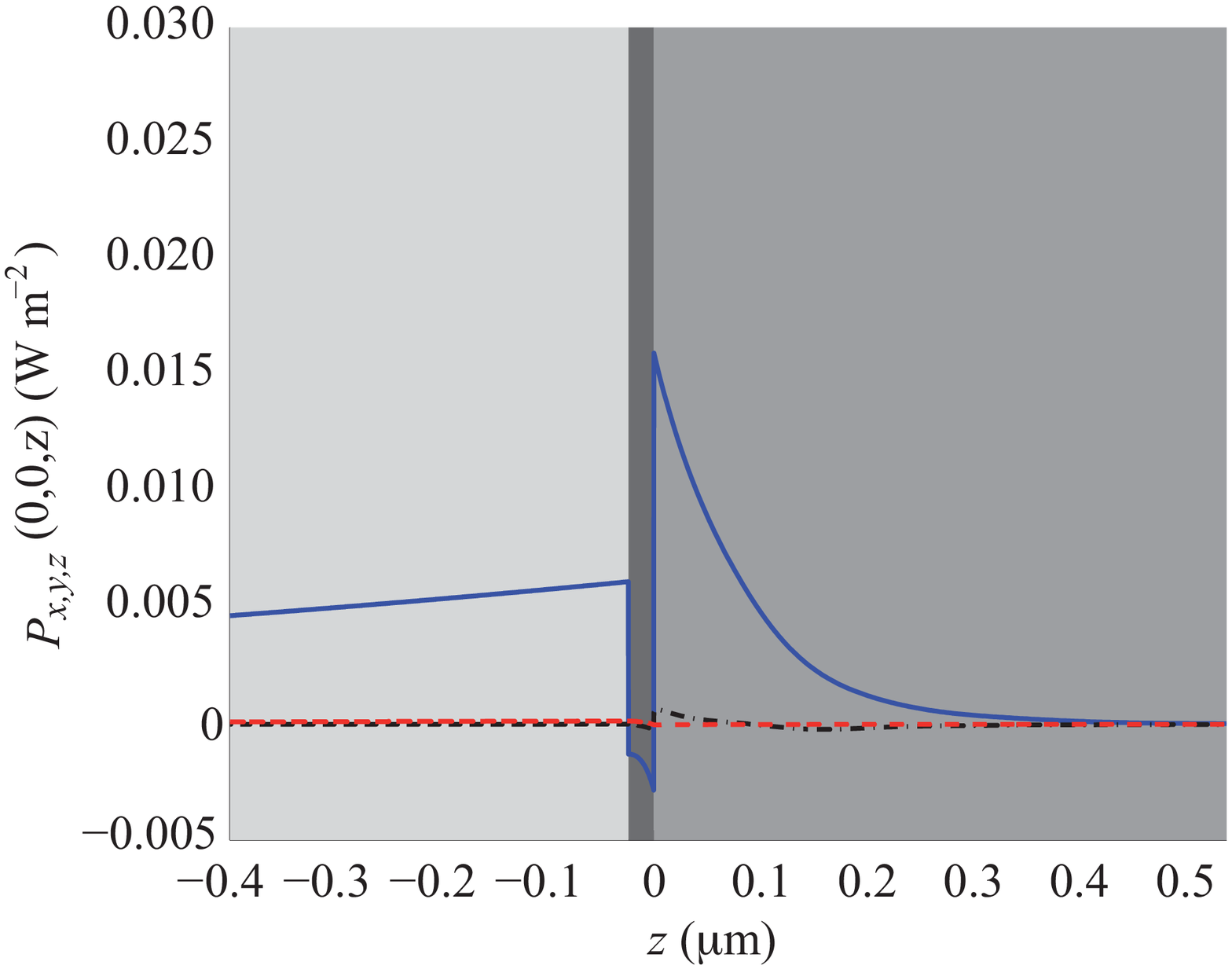}\hspace{10pt}
(d) \includegraphics[width=0.4\linewidth]{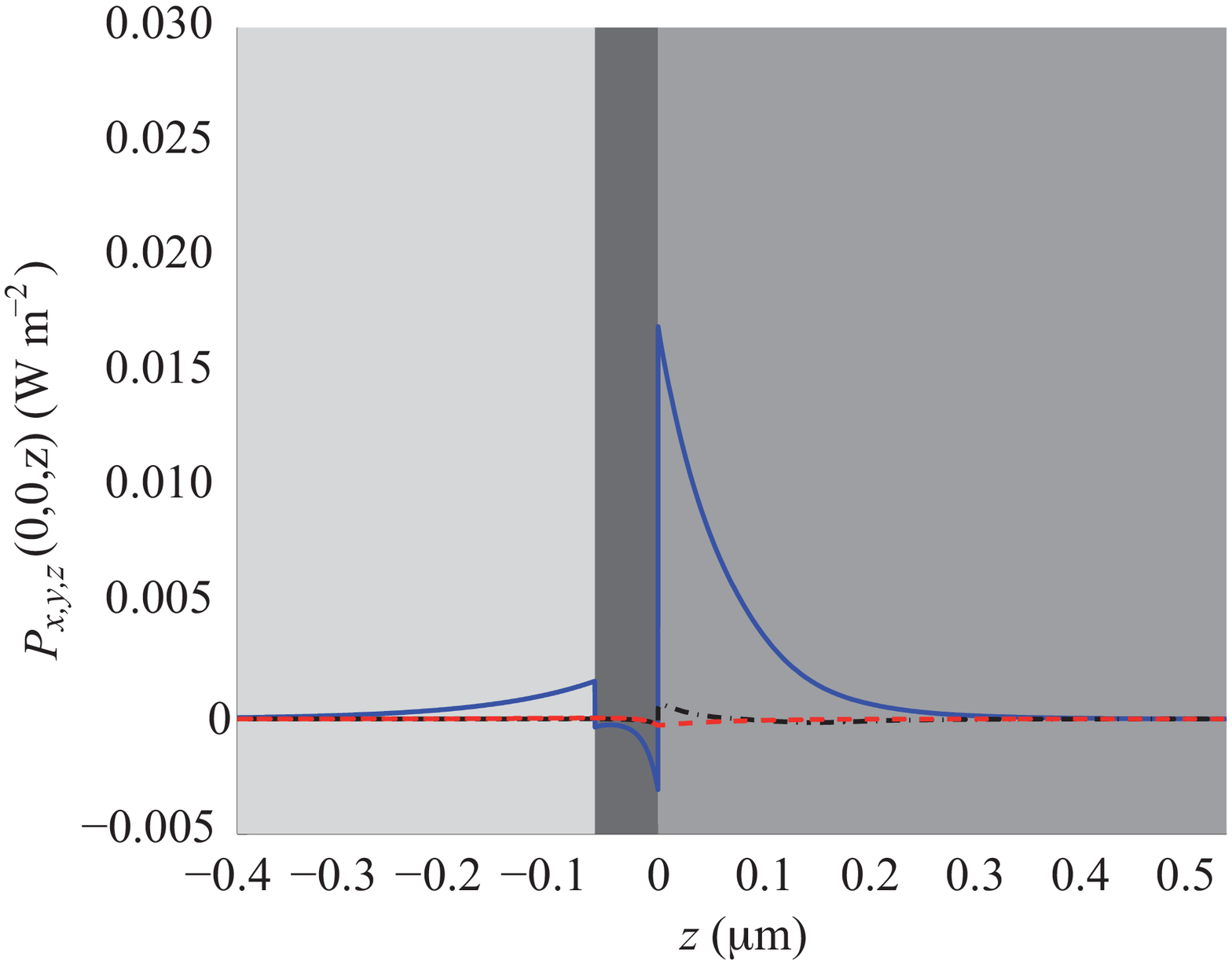}\\
(e) \includegraphics[width=0.4\linewidth]{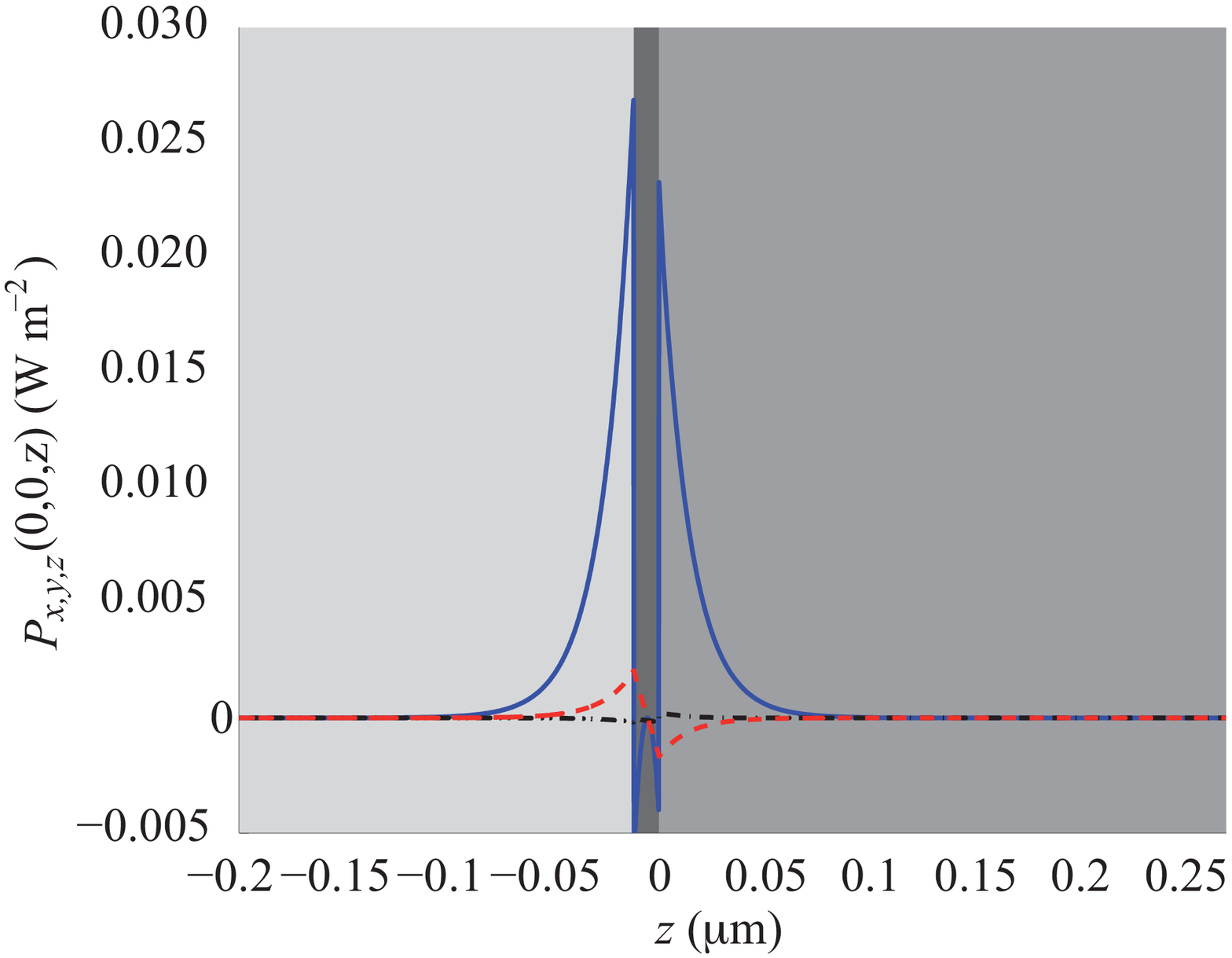}\hspace{10pt}
(f) \includegraphics[width=0.4\linewidth]{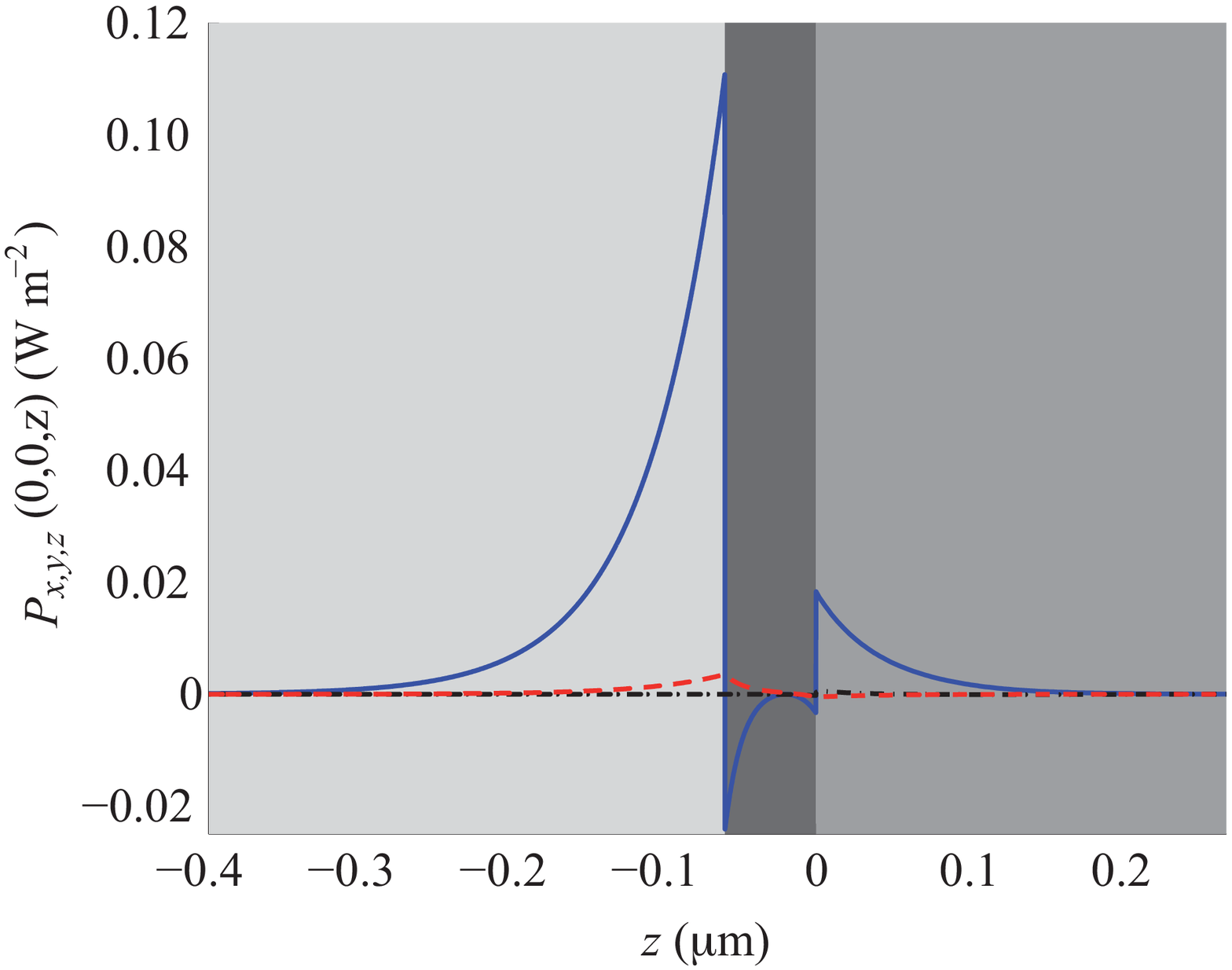}
 \end{tabular}
 \end{center}
\caption{
Variations of $P_x(0,0,z)$ (blue solid lines), $P_y(0,0,z)$ (black dotted-dashed lines), and $P_z(0,0,z)$ (red dashed lines) with respect to $z$ for the compound SPP waves when
$\psi=0^\circ$ and
(a) $L=48$~nm or (b) $L=60$~nm for  branch 1;
(c) $L=24$~nm or (d) $L=60$~nm for  branch 2; and
(e) $L=12$~nm or (f) $L=60$~nm for  branch 3. All field phasors have been normalized so that numerical comparisons are possible.
 \label{fig:SupersetPa}}
 \end{figure*}

The foregoing conclusions are confirmed by the plots of the spatial profiles of the Cartesian components of the time-averaged Poynting vector $ \textbf{P}\left(x,y,z \right) $. 
\fref{fig:SupersetPa} shows the spatial variations of $P_x(0,0,z)$, $P_y(0,0,z)$ and $P_z(0,0,z)$  for representative compound SPP waves on the three solution branches,
when when $\psi=0^\circ$. From the figure it is evident that  {the compound SPP wave belonging to   branch $ 1 $ propagates}
bound predominantly to the metal/SCM interface with its energy confined mostly to  several periods of the SCM close
to the metal. The compound SPP wave on  branch $ 2 $ is bound to the metal/SCM interface only when $L$ is large enough, and most of its energy is contained
in the first period of the SCM.  As $L$ decreases, the wave becomes increasingly bound also to the metal/HID interface, thereby indicating a significant coupling between the 
two metal/dielectric interfaces. The compound SPP wave belonging to  branch $ 3 $ propagates bound predominantly to the metal/HID interface when $L=60$~nm with a power density confined almost totally in the HID material. The coupling between the two interfaces is clearly evident when $ L =12$~nm, the wave then propagating bound to both
metal/dielectric interfaces and its energy  distributed almost equally in both the HID material and the dielectric SCM.

\begin{table*} 
\caption{\bf Values of $ \tq$ computed for compound SPP waves belonging to the three solution branches when $\psi=0^\circ$ and
the metal layer has either a thickness $L=60$~nm or $ L \simeq L_{th} $, where $ L_{th} $ is the minimum thickness for which a solution 
on a specific branch was found.
Solutions obtained for SPP waves guided by either the metal/SCM interface alone
or the metal/HID interface alone are also provided.
\label{tab:Ssol}}
\begin{center} 
\begin{tabular}{c|ccc}
\hline
\cline{2-4}
\rule[-1ex]{0pt}{3.5ex} Branch $\rightarrow$ & $1$ & $2$ & $3$\\
\hline\hline
\rule[-1ex]{0pt}{3.5ex} $L\simeq L_{th}$ & $1.1356+i 0.0005$ & $1.7785 + i 0.0008 $ & $3.9514 + i 0.2679$ \\
\rule[-1ex]{0pt}{3.5ex} $L=60$~nm & $1.1344 + i0.0018$ & $1.8436 + i 0.0133$ & $2.0417+ i 0.0321$ \\
\hline\hline
$\tq_{met/SCM}$ &  $1.1340 + i 0.0025$ & $1.8584+ i0.0181$ & - \\
\hline\hline
$\tq_{met/HID}$ & - & - &{$2.0100 + i0.0230 $}\\
\hline\hline
\end{tabular}
\end{center}
\end{table*}

\subsection{Effect of the relative permittivity of the HID material}\label{variablepsd}
In Sec.~\ref{variableL} we showed that the number of the compound SPP waves guided jointly by the two metal/dielectric interfaces reduces
(and even vanishes) as the thickness $L$ of the metal layer decreases; moreover, no compound SPP wave exists when $ L <\delta_m/2.2$. In order to investigate the influence of the relative permittivity of the HID material on the number of  compound SPP waves, we fixed 
$\psi=0^\circ$ and $L = 25$~nm
(i.e., $L$ is slightly smaller than $\delta_m$), and varied $\eps_d$   from $1$ to $4$.

\begin{figure}
 \begin{center}
 \begin{tabular}{c}
\includegraphics[width=0.9\linewidth]{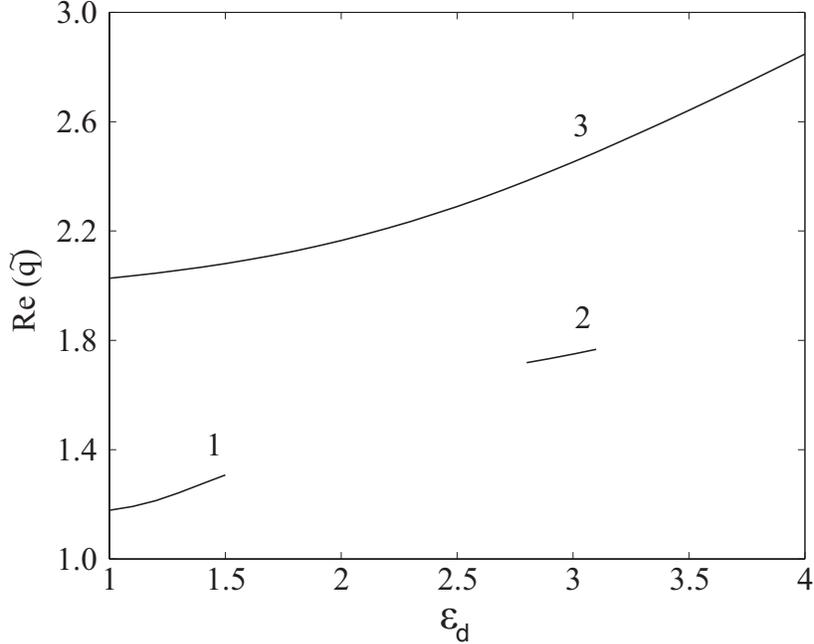}
\end{tabular}
 \end{center}
\caption{Variation of ${\rm Re}(\tq)$ with the relative permittivity $\eps_d$ of the HID material when $L=25$~nm and $\psi=0^\circ$. 
\label{fig:Req}}
 \end{figure}

\begin{figure}
 \begin{center}
 \begin{tabular}{c}
 \includegraphics[width=0.9\linewidth]{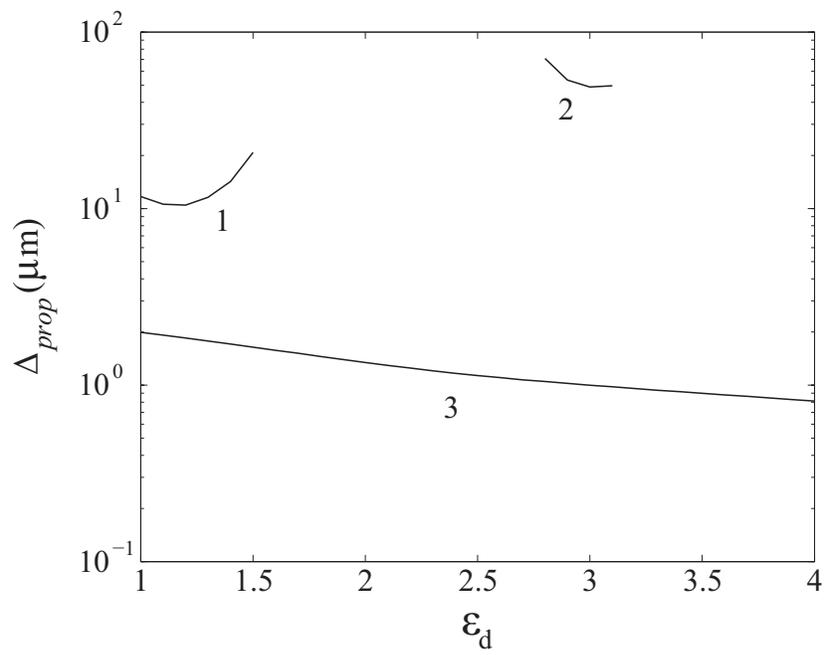}
 \end{tabular}
 \end{center}
\caption{Variation of the propagation distance $\propdist$ with the relative permittivity
$\eps_d$ of the HID material when $L=25$~nm and $\psi=0^\circ$. 
 \label{fig:Deltaprop}}
 \end{figure}

Figures \ref{fig:Req} and \ref{fig:Deltaprop}, respectively, show plots of ${\rm Re}(\tq)$ and $\propdist$ as functions of $\eps_d$. The solutions are organized in three branches labeled $1$ to $3$. Two solutions exist in two distinct $\eps_d$-intervals: (i) $ 1 \leq \eps_d \lesssim 1.5 $ and 
(ii) $ 2.8 \lesssim \eps_d \lesssim 3.1 $. In the remaining parts of the range $1\leq\eps_d\leq4$,  only 
one  compound SPP wave exists. As $ \eps_d $ increases, the phase speed $ \vph $ increases on all three branches, the increase being almost
linear on branches 1 and 2. As $\eps_d$ increases, the propagation distance $\propdist$ first decreases and then increases
on branches 1 and 2, but decreases almost linearly on branch 3.
 

\begin{figure}
 \begin{center}
 \begin{tabular}{c}
 (a) \includegraphics[width=0.9\linewidth]{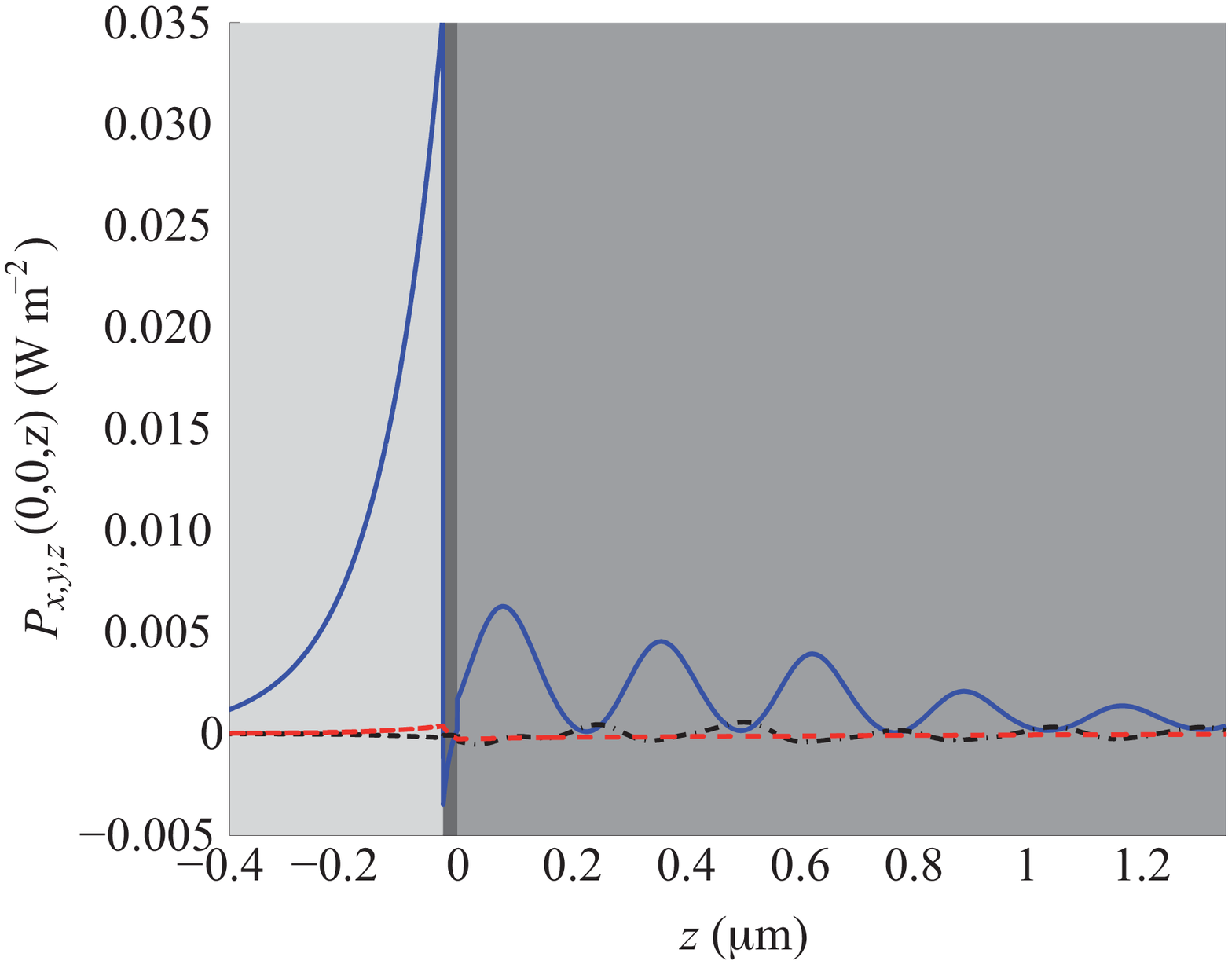}\\
 (b) \includegraphics[width=0.9\linewidth]{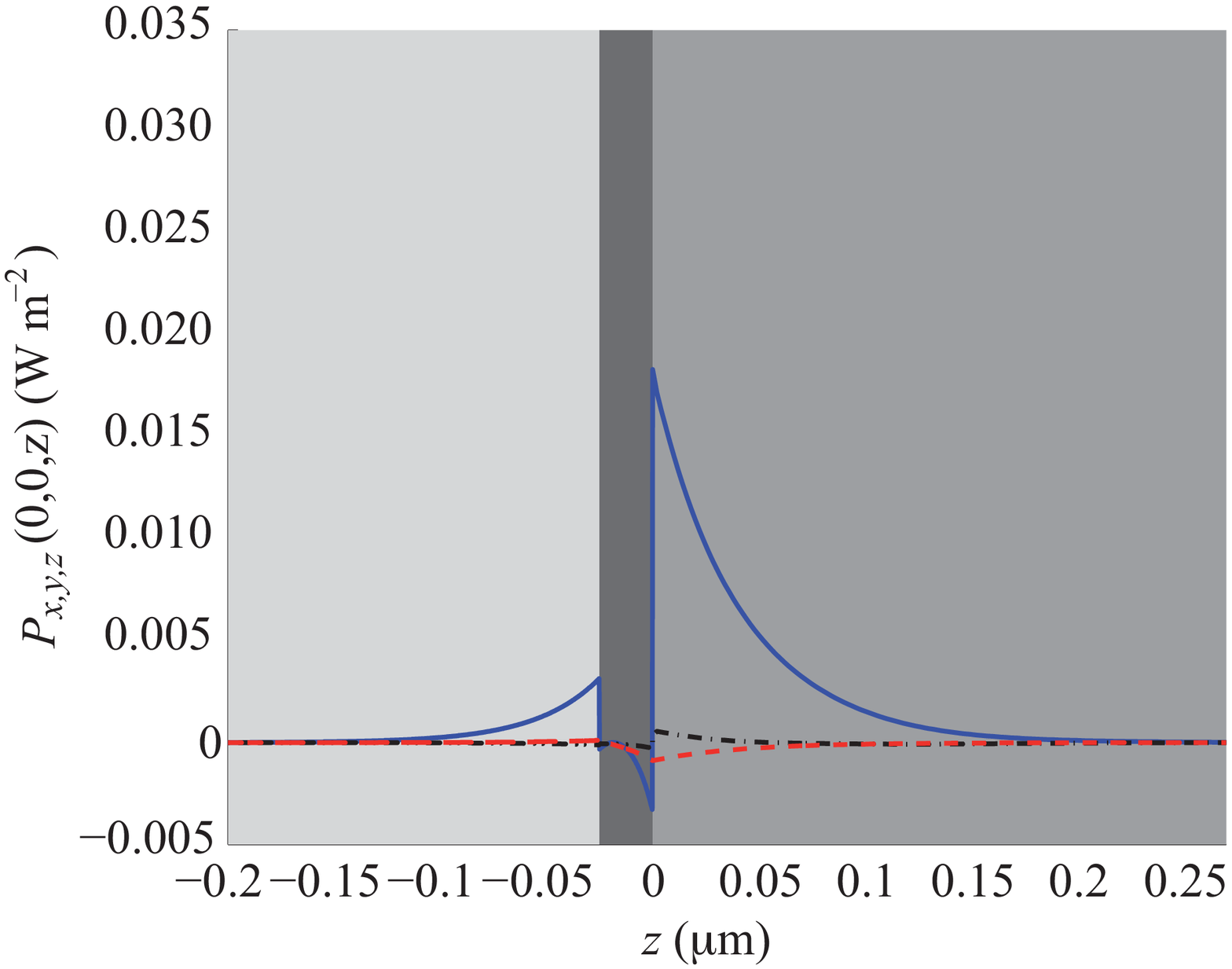}
 \end{tabular}
 \end{center}
 \caption{Variations of $P_x(0,0,z)$ (blue solid lines), $P_y(0,0,z)$ (black dotted-dashed lines), and $P_z(0,0,z)$ (red dashed lines) with respect to $z$ for the compound SPP wave on (a) branch 1 or (b) branch 3 when $\eps_d=1.5$, $L=25$~nm, and $\psi=0^\circ$.
 \label{fig:CSSPppol}}
 \end{figure}

Figure \ref{fig:CSSPppol}  shows the spatial profiles of the Cartesian components of the time-averaged Poynting vector $\#P(0,0,z)$ for  compound SPP waves belonging to the branches $ 1 $ and $ 3 $ when $ \eps_d =1.5$. Likewise,
Fig.~\ref{fig:CSSPppol}   shows the spatial profiles of the Cartesian components  $\#P(0,0,z)$ for  compound SPP waves belonging to the branches   $ 2 $ and $ 3 $ when $ \eps_d =2.8$. Due to the significant coupling between the two dielectric/metal interfaces, significant fractions of the energy of the compound SPP waves reside in the both the HID material and the
dielectric SCM. 
When $ \eps_d=1.5 $, most of the energy resides in the HID material for the wave on  branch $ 1 $ and in the dielectric SCM for the wave on branch $ 3 $. When $ \eps_d=2.8 $, the energy of the wave is distributed almost equally in both dielectric materials.
 \begin{figure}
 \begin{center}
 \begin{tabular}{c}
 (a) \includegraphics[width=0.9\linewidth]{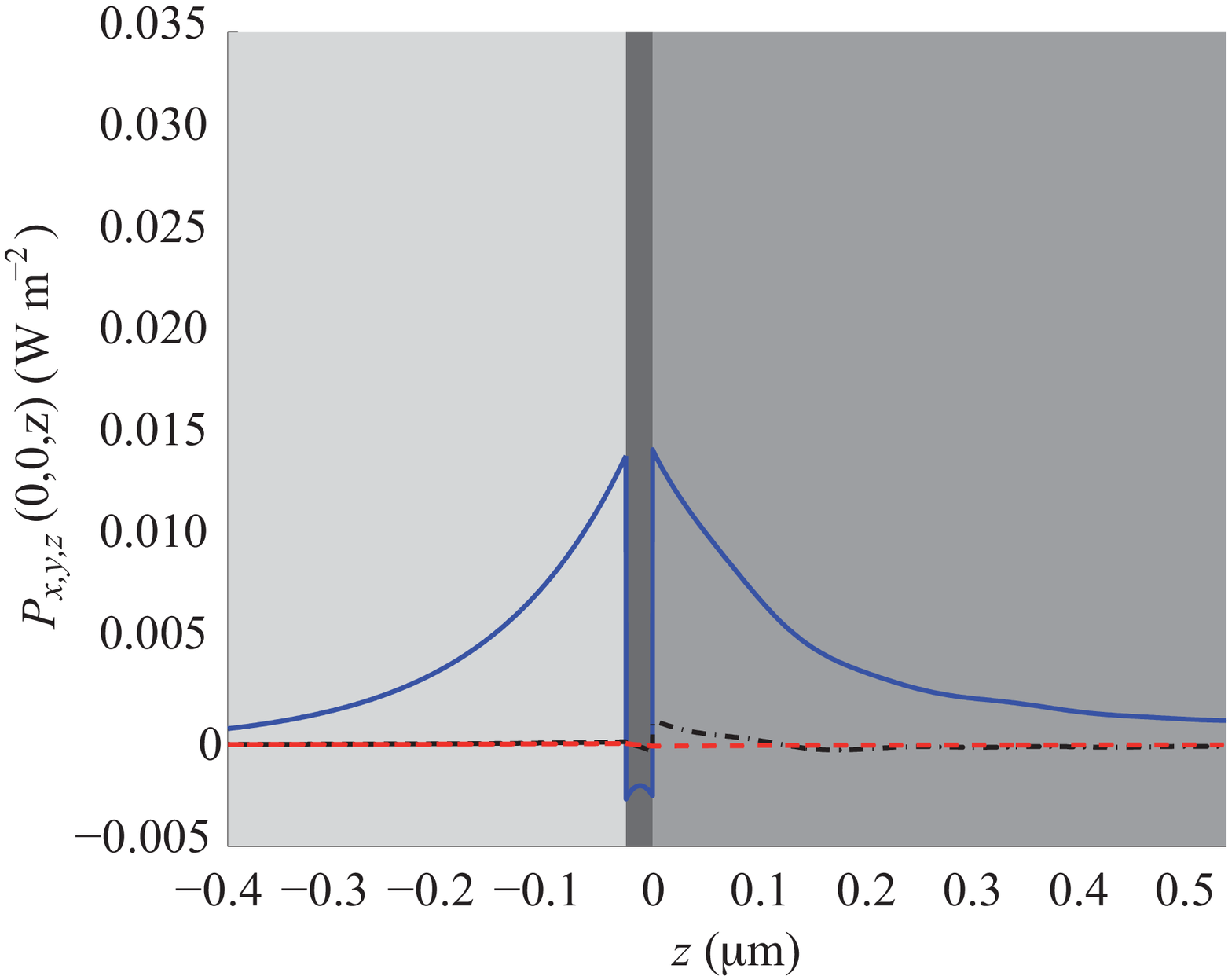}\\
 (b) \includegraphics[width=0.9\linewidth]{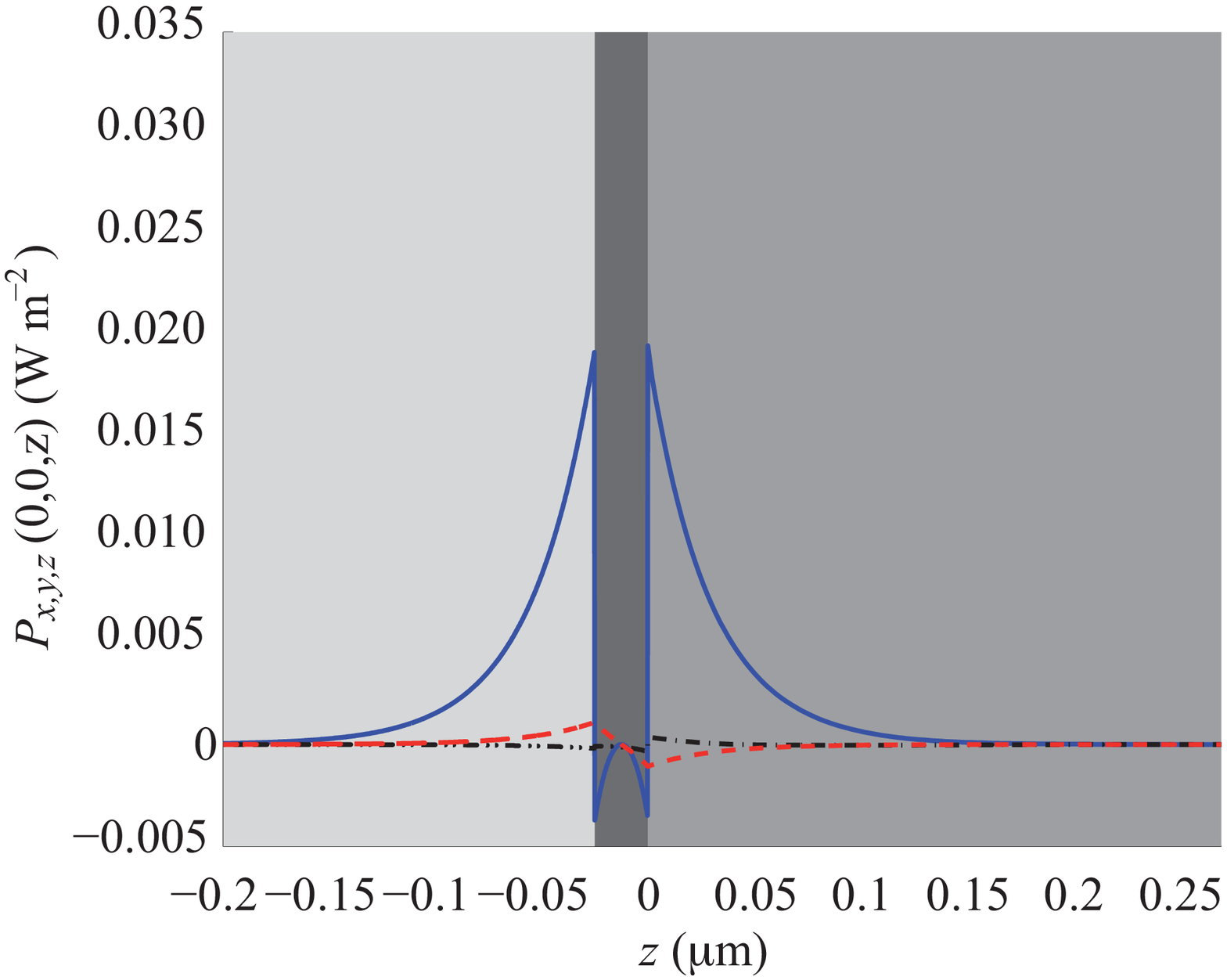}
 \end{tabular}
 \end{center}
 \caption{Variations of $P_x(0,0,z)$ (blue solid lines), $P_y(0,0,z)$ (black dotted-dashed lines), and $P_z(0,0,z)$ (red dashed lines) with respect to $z$ for the compound SPP wave on (a) branch 2 or (b) branch 3 when $\eps_d=2.8$, $L=25$~nm and $\psi=0^\circ$. 
 \label{fig:CSSPspol}}
 \end{figure} 

\subsection{Influence of the transverse anisotropy of the SCM}\label{sec:psi}
So far in this paper, we have reported results for $\psi=0^\circ$.
Since  an SCM is not isotropic in the $xy$ plane, the choice of the direction of propagation as quantified
through $\psi$ should have an effect. In order to analyze the effect of the anisotropy of the
dielectric SCM, we fixed $\eps_d=3.1634$ (SF11 glass) and varied $\psi\in\left[-90^\circ,90^\circ\right]$ for $L\in\left\{25,60\right\}$~nm. Due to the symmetry of the 
$\=\eps_{SCM}\left(z\right)$ in the $xy$ plane, the solutions for $\psi\pm180^\circ$ are the same for $\psi$.

\begin{figure}
 \begin{center}
 \begin{tabular}{c}
(a) \includegraphics[width=0.8\linewidth]{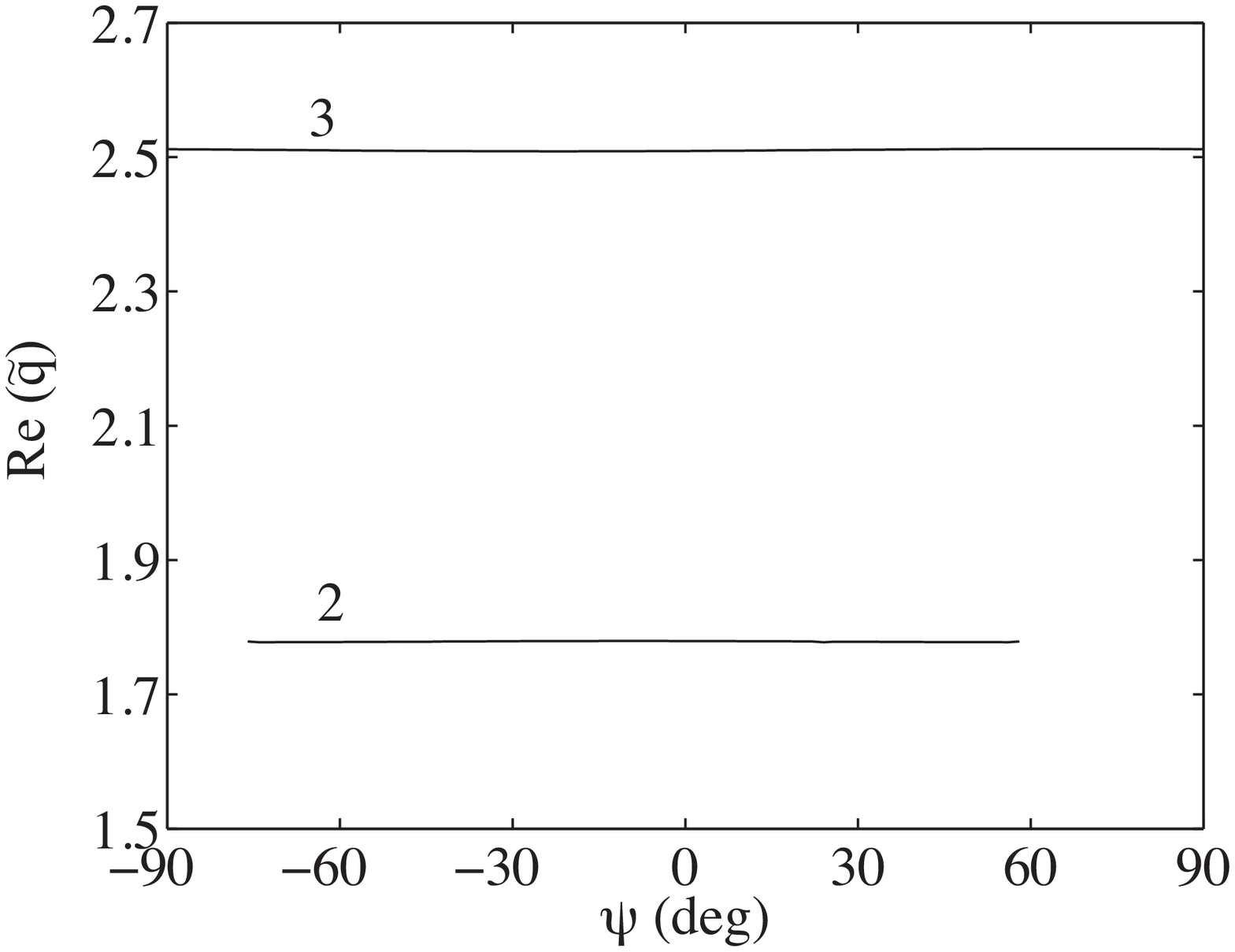}\\[20pt]
(b) \includegraphics[width=0.8\linewidth]{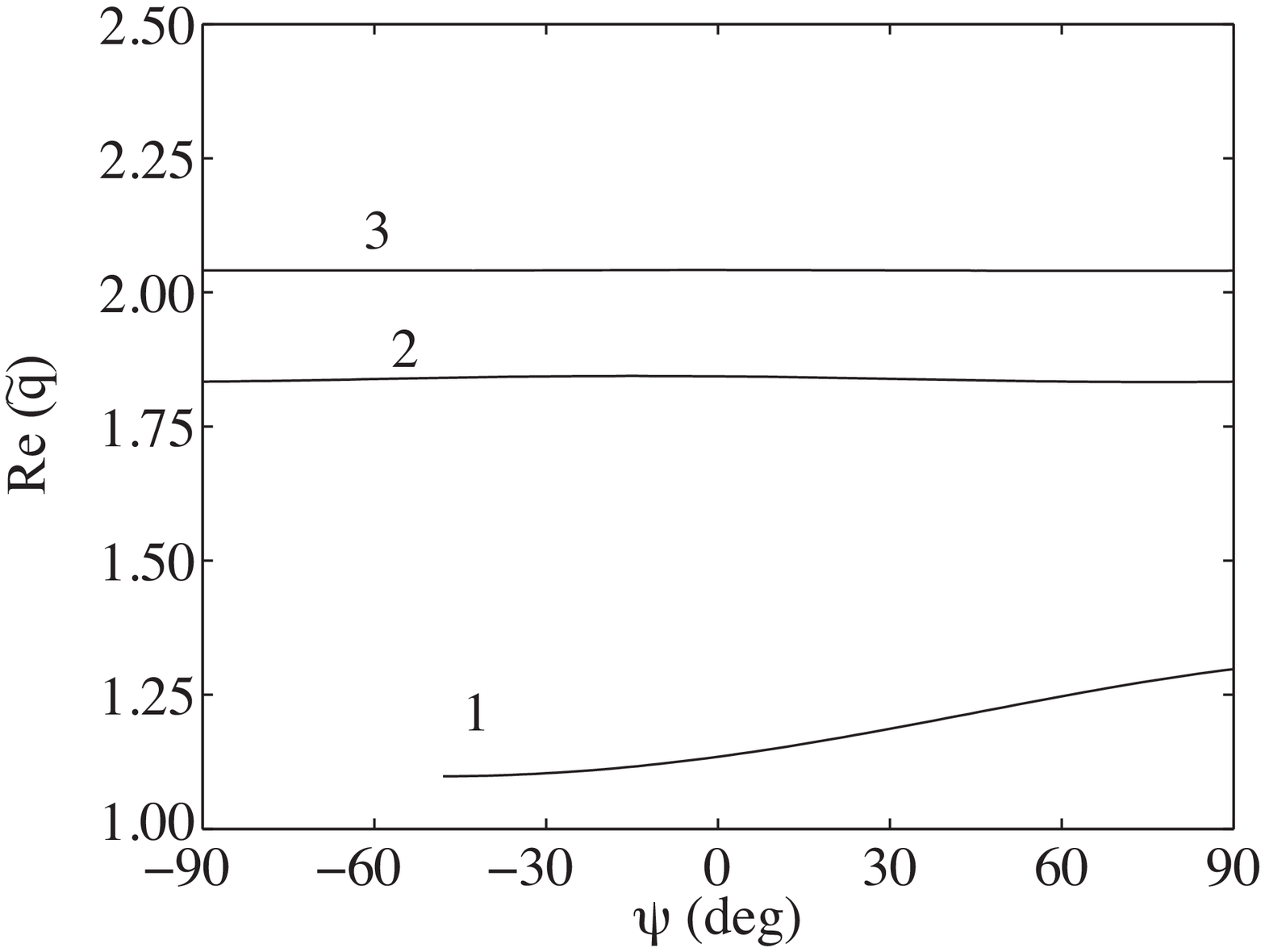}
\end{tabular}
 \end{center}
\caption{Variation of ${\rm Re}(\tq)$ with the angle $\psi$  when  $\eps_d=3.1634$ and
(a) $L=25$ nm or (b) $L=60$ nm. 
\label{fig:ReqPsi}}
 \end{figure}

\begin{figure}
 \begin{center}
 \begin{tabular}{c}
(a) \includegraphics[width=0.8\linewidth]{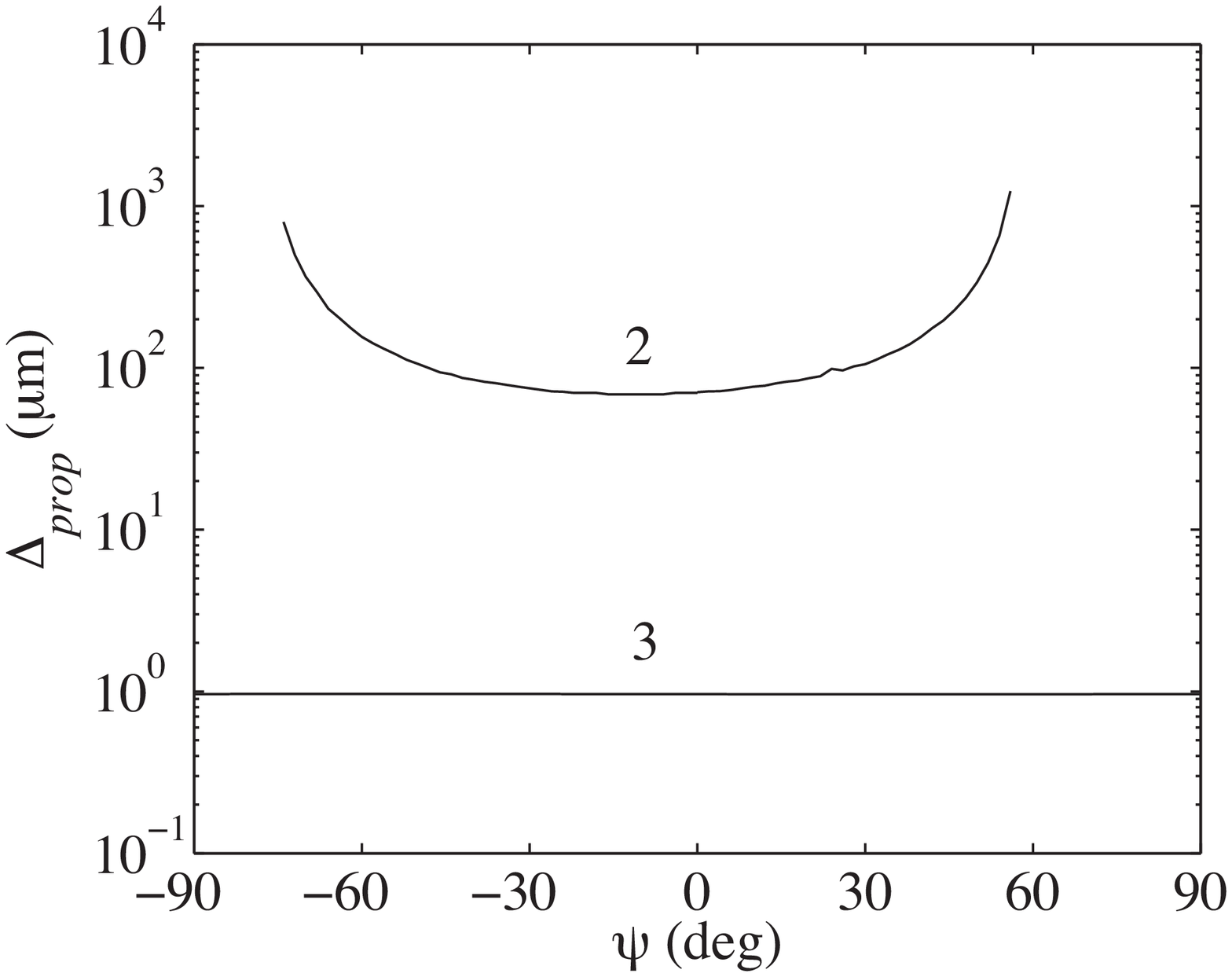}\\[20pt]
(b) \includegraphics[width=0.8\linewidth]{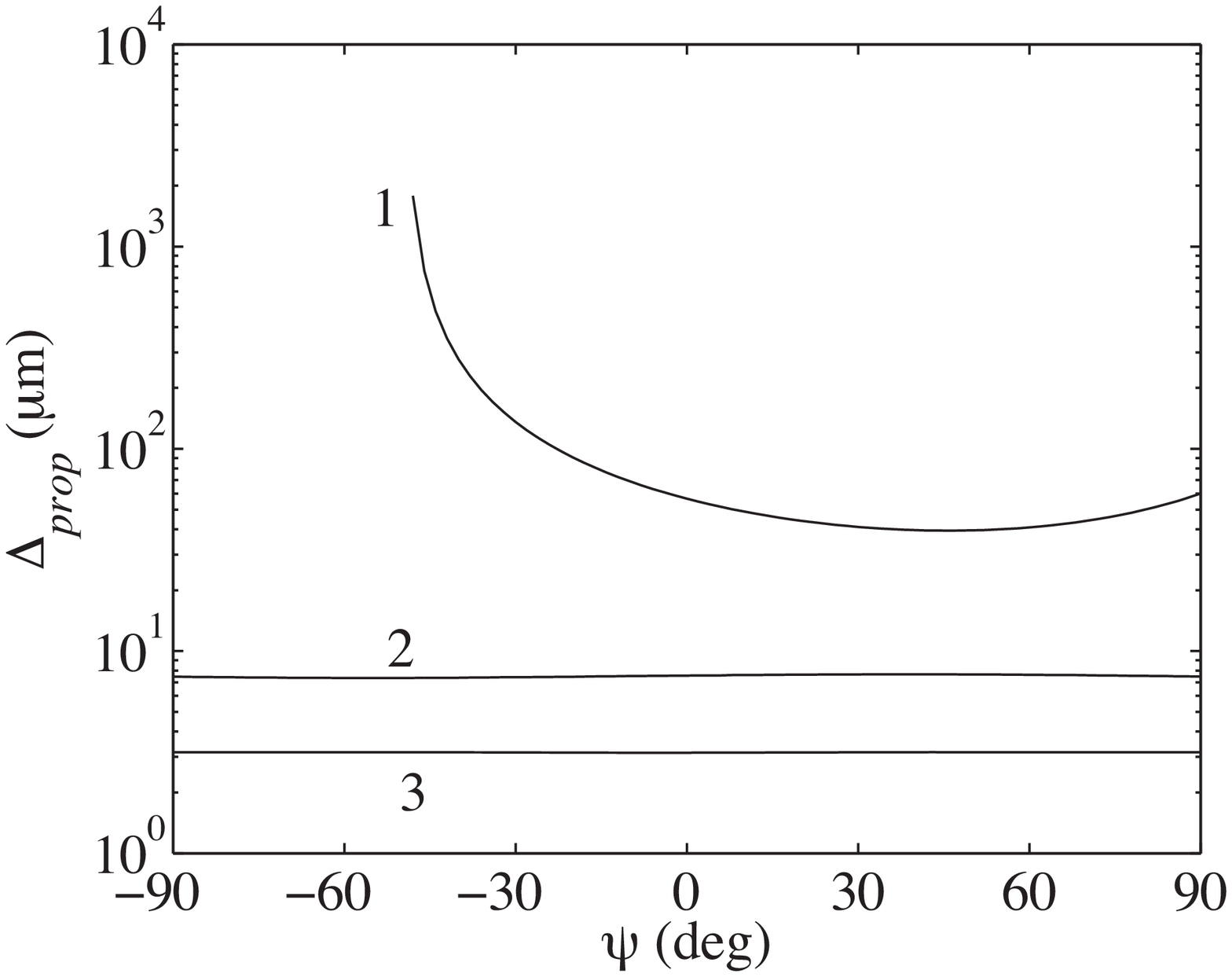}
 \end{tabular}
 \end{center}
\caption{Variation of the propagation distance $\propdist$ with the angle $\psi$  when  $\eps_d=3.1634$ and (a) $L=25$ nm or (b) $L=60$ nm. 
 \label{fig:DeltapropPsi}}
 \end{figure}

Figures \ref{fig:ReqPsi} and \ref{fig:DeltapropPsi}, respectively, show plots of ${\rm Re}(\tq)$ and $\propdist$ as functions of $\psi$ for $L\in\left\{25,60\right\}$~nm. In these figures, the solution branches are labeled to be consistent with the data presented in Figs.~\ref{fig:ReqL} and \ref{fig:DeltapropL} for $\psi=0^\circ$.

For the thinner metal layer ($L = 25$~nm),  branch~1 does not exist for any $\psi$. Whereas
branch~2  is confined to $\psi\in[-76^\circ,58^\circ]$, branch~3 spans the entire range $\psi\in\left[-90^\circ,90^\circ\right]$. On both branches, Re$\tond{\tilde{q}}$ is almost constant, which
implies that $\vph$ is very weakly dependent on $\psi$. However, whereas
 $\propdist$  is almost independent of $\psi$ on branch~3, it varies by a factor of 10 on 
 branch~2 with a minimum at $\psi=-10^\circ$.
   
Solution branches 1--3 exist for   $L=60$~nm. Two of those branches (labeled 2 and 3) span the entire range of $\psi$ while one (labeled 1) is confined to $\psi\in\quadr{-48^\circ,90^\circ}$.   Both Re$\tond{\tilde{q}}$ (and, therefore, $\vph$) and $\propdist$ depend on $\psi$ very weakly on branches 2 and 3. In contrast, Re$\tond{\tilde{q}}$  increases and $\vph$ decreases almost monotonically  as $\psi$ increases, but $\propdist$  decreases up to $\psi=46^\circ$ where it reaches its minimum and then increases.

Thus, we  conclude that  some compound SPP waves is not affected greatly by the direction of propagation but    others are, depending on metal thickness. A device containing the HID/metal/SCM structure could be rotated about the $z$ axis for optimal performance.

\section{Concluding remarks}\label{sec:concl}
 
The boundary-value problem for the propagation of
compound surface plasmon-polariton waves guided by a thin metal layer sandwiched between  a homogeneous isotropic dielectric   material  and a dielectric structurally chiral material was formulated and solved numerically. 

Multiple compound SPP waves, differing in phase speed, attenuation rate and field profile, were found to exist  at a fixed frequency (or free-space wavelength, equivalently) only if the thickness of the metal layer  is sufficiently large.  
When the metal thickness is small, the two metal/dielectric interfaces  couple to each  other,  thereby resulting in one or more compound SPP waves that propagate bound to both interfaces with energy lying in both the HID material and the dielectric  SCM.
As the metal thickness increases,  the coupling between the two dielectric/metal interfaces  weakens. Eventually, each compound SPP wave is bound predominantly to just one of the two metal/dielectric interfaces with its energy residing mostly in the dielectric partner.  If the thickness of the metal layer is greater than the skin depth, the attributes of the compound SPP waves approach those of the SPP waves guided by a metal/dielectric interface alone. The propagation characteristics of some of the compound SPP waves are less  affected by the transverse anisotropy of the dielectric SCM than of others.

Thus, the propagation characteristics can be tailored by appropriate choices of the metal thickness, the relative permittivity of the HID material, and the direction of propagation in the transverse plane, once the dielectric SCM has been chosen.

Finally, the multiplicity and   the number of compound SPP waves depend on the relative permittivity of the HID material. This feature could be useful for optical sensing applications.

\end{document}